\documentclass[lineno]{jfm}

\usepackage{graphicx}
\usepackage{epstopdf,epsfig}
\usepackage{newtxtext}
\usepackage{newtxmath}
\usepackage{natbib}

\newcommand{\RomanNumeralCaps}[1]
\nolinenumbers

\usepackage{bm}%
\usepackage[colorlinks,
linkcolor=blue,
anchorcolor=blue,
citecolor=blue
]{hyperref}
\usepackage{subfigure}
\usepackage{multirow}
\usepackage{natbib}
\usepackage{float}
\usepackage{makecell}
\usepackage{amsmath}

\usepackage{color}

\shorttitle{Inclination angle of wall-attached eddies}
\shortauthor{C. Cheng, W. Shyy \&  L. Fu}

\title{Streamwise inclination angle of wall-attached eddies in turbulent channel flows }

\author{
Cheng Cheng\aff{1}, Wei Shyy\aff{1},
Lin Fu\aff{1}$^,$\aff{2}$^,$\aff{3}$^,$\corresp{\email{linfu@ust.hk}}
}

\affiliation{
\aff{1}Department of Mechanical and Aerospace Engineering, The Hong Kong University of Science and Technology, Clear Water Bay, Kowloon, Hong Kong
\aff{2}Department of Mathematics, The Hong Kong University of Science and Technology, Clear Water Bay, Kowloon, Hong Kong
\aff{3} Center for Ocean Research in Hong Kong and Macau (CORE), The Hong Kong University of Science and Technology, Clear Water Bay,
Kowloon, Hong Kong
}

\begin{document}
\maketitle
\nolinenumbers

\begin{abstract}
We develop a new methodology to assess the streamwise inclination angles (SIAs) of the wall-attached eddies populating the logarithmic region with a given wall-normal height. To remove the influences originating from other scales on the SIA estimated via two-point correlation, the footprints of the targeted eddies in the vicinity of the wall and the corresponding streamwise velocity fluctuations carried by them are isolated simultaneously, by coupling the  spectral stochastic estimation with the attached-eddy hypothesis. Datasets produced with direct numerical simulations spanning $Re_{\tau} \sim O(10^2)-O(10^3)$ are dissected to study the Reynolds-number effect. The present results show, for the first time, that the SIAs of attached eddies are Reynolds-number dependent in low and medium Reynolds numbers and tend to saturate at $45^{\circ}$ as the Reynolds number increases. The mean SIA reported by vast previous experimental studies are demonstrated to be the outcomes of the additive effect contributed by multi-scale attached eddies. These findings clarify the long-term debate and perfect the picture of the attached-eddy model.

\end{abstract}


{\bf MSC Codes }  {\it(Optional)} Please enter your MSC Codes here

\section{Introduction}
\begin{figure} 
\centering 
\subfigure{ 
	\label{fig:AEH:a} 
	\includegraphics[width=3.5in]{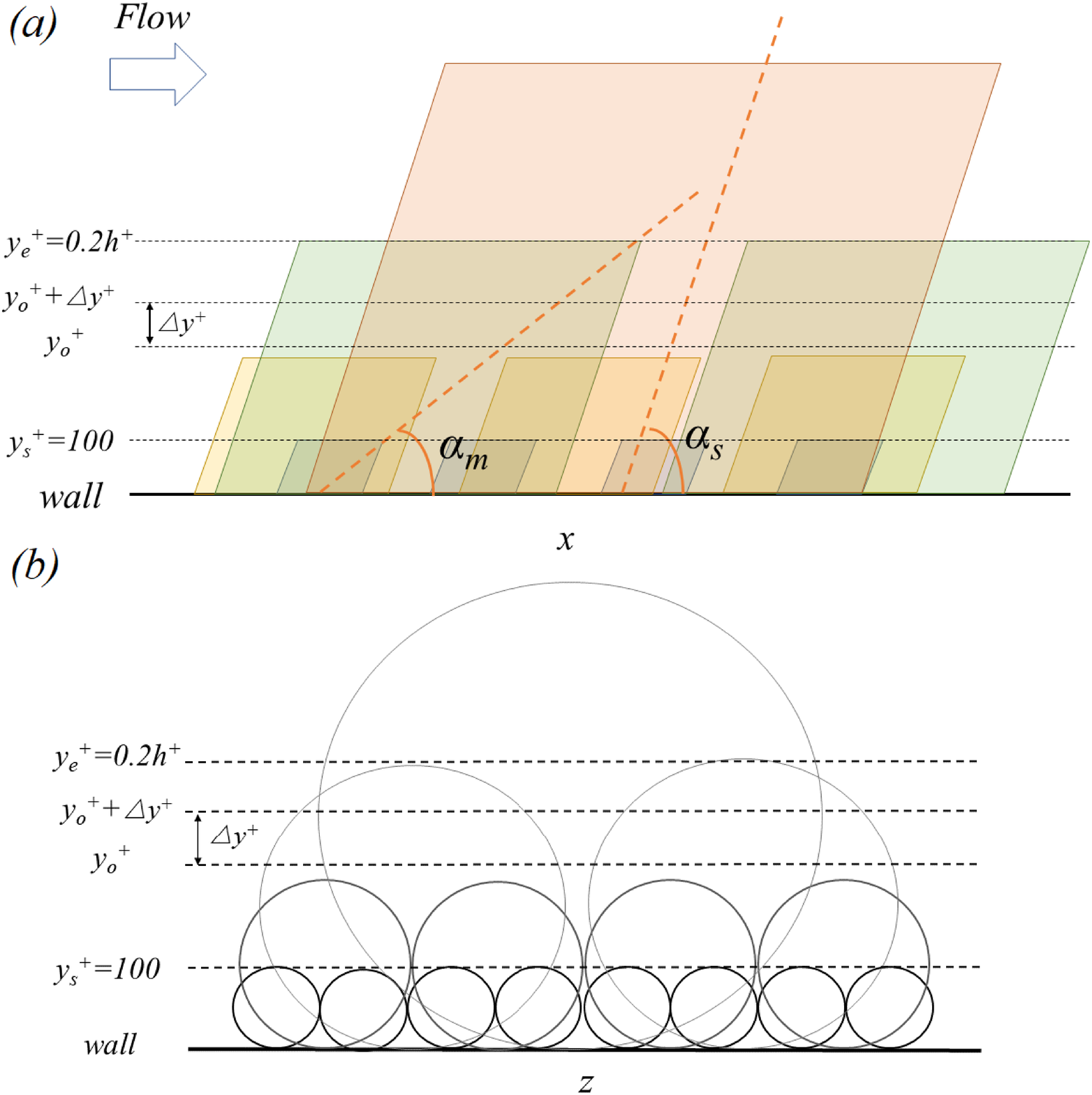} 
}
\caption{A schematic of the attached-eddy model: ($a$) $x$-$y$ plane and ($b$) $y$-$z$ plane view. Each parallelogram in ($a$) and circle in ($b$) represents an individual attached eddy.
	Here, $x$, $y$ and $z$ denote the streamwise, wall-normal, and spanwise directions, respectively. $y_s^+$ (100) and $y_e^+$ ($0.2h^+$) denote the lower and upper bound of the logarithmic region, respectively \citep{Jimenez2018,Baars2020a,Wang2021}. $y_o^+$ is the outer reference height. $\Delta y^+$ is the local grid spacing along the wall-normal direction. $\alpha_m$ and $\alpha_s$ are the mean and individual SIA of attached eddies, respectively. These two figures are merely conceptual sketches, and the eddy population density is not in accordance with that of \cite{Perry1982}.}
\label{fig:AEH} 
\end{figure}

It is generally recognized that the high-Reynolds number wall-bounded turbulence is filled with coherent motions of disparate scales, which are responsible for the energy transfer and the fluctuation generation of turbulence. Till now, the most elegant conceptual model describing these energy-containing motions is the attached-eddy model \citep{Townsend1976,Perry1982}. It hypothesizes that the logarithmic region is occupied by an array of randomly-distributed and self-similar energy-containing motions (or eddies) with their roots attached to the near-wall region (see Fig.~\ref{fig:AEH}). During the recent decades, a growing body of evidence that supports the attached-eddy hypothesis has emerged rapidly, e.g., \cite{Hwang2015}, \cite{Hwang2018}, \cite{Cheng2020}, \cite{Hwang2020}, to name a few. The reader is referred to a recent review work by \cite{Marusic2019} for more details. Throughout the paper, the terms `eddy' and `motion' are exchangeable. It should be noted that the terms of `wall-attached motions' and `wall-attached eddies' used in the present study do not only refer to the self-similar eddies in the logarithmic region, but also the very-large-scale motions (VLSMs) or superstructures, as some recent studies have shown that VLSMs are also wall-attached, despite that their physical characteristics do not match the attached-eddy model \citep{Hwang2018,yoon2020}.

Previous studies have established that the energy-containing eddies populating the logarithmic and outer regions bear characteristic SIAs due to the mean shear (see Fig.~\ref{fig:AEH}($a$)). As early as the 1970s, \cite{Kovasznay1970} found that the large-scale structures in the outer intermittent region of a turbulent boundary layer have a moderate tilt in the streamwise direction. On the other hand, for the eddies in the logarithmic region of wall turbulence, the wall-attached $\Lambda-$vortex was used by \cite{Perry1982} to illustrate them. According to \cite{Adrian2000}, these $\Lambda-$vortexes are apt to cluster along the flow direction and form an integral whole (generally called as vortex packets). Further observations in channel flows \citep{Christensen2001} demonstrated that the heads of $\Lambda-$vortexes among the vortex packets tend to slope away from the wall in a statistical sense, with SIAs between $12^{\circ}$ and $13^{\circ}$. Most additional studies have shown a similar result, and it is widely accepted that the approximate SIAs of eddies are in the range of $10^{\circ}$ to $16^{\circ}$  \citep{Boppe1999,Christensen2001,carper2004,Marusic2007,Baars2016}. Besides, the SIA is also found to be Reynolds-number independent \citep{Marusic2007}.

However, the SIA estimated by experimentalists using the traditional statistical approach is the mean structure angle indeed \citep{Marusic2007,Deshpande2019}. The common procedure to obtain the SIA is based on the calculation of the cross correlation between the streamwise wall-shear stress fluctuation ($\tau_x'$) and the streamwise velocity fluctuation ($u'$) at a wall-normal position in the log region ($y_o$). The cross correlation can be expressed as
\begin{equation}\label{corr}
R_{\tau_x' u'}(\Delta x)=\frac{\langle\tau_x'(x) u'(x+\Delta x,y_o)\rangle}{\sqrt{\left\langle\tau_x'^{2}\right\rangle\left\langle u^{'2}\right\rangle}},
\end{equation}
where $<\cdot>$ represents the ensemble spatial average, and $\Delta x$ the streamwise delay. The SIA can be estimated by 
\begin{equation}\label{mangle}
\alpha_m=\arctan(\frac{y_o}{\Delta x_{p}}),
\end{equation}
where $\Delta x_{p}$ denotes the streamwise delay corresponding to the peak in $R_{\tau_x' u'}$. Considering that an array of wall-attached eddies with distinct wall-normal heights can simultaneously convect past the reference position $y_o$, $\alpha_m$ in Eq.~(\ref{mangle}) should be regarded as the mean angle of these eddies. Hence, the subscript `m' in Eq.~(\ref{mangle}) refers to `mean'. 

To estimate the SIAs of the largest wall-attached eddies, \cite{Deshpande2019} introduced a spanwise offset between the near-wall and logarithmic probes to isolate these wall-attached motions in the log region. They found that their SIAs are approximately $45^{\circ}$. This observation is consistent with several theoretical analyses. For example, \cite{Moin1985} and \cite{Perry1992} proposed that for the flows with two-dimensional mean flows, the characteristic angles of the energy-containing eddies should follow the direction of the principal rate of mean strain. More specifically, their SIAs should be $45^{\circ}$ for a zero-pressure-gradient turbulent boundary layer \citep{Perry1992}. \cite{Marusic2001} found that the mean SIA of the induced turbulence field by attached eddies is akin to the experimental measurements, if the hierarchical attached eddies tilt away from the wall with individual SIA being $45^{\circ}$ and organize like the vortex packets observed in numerical and laboratory experiments.

Reviewing the work of predecessors, it can be found the SIAs of attached eddies at a given length scale are ambiguous. Traditional measurements are only applicable for the assessment of the mean SIA \citep{brown1977,Boppe1999,Marusic2007}. Moreover, the technique adopted by \cite{Deshpande2019} can only isolate the largest wall-attached motions in the logarithmic region. Considering the characteristic scale of an individual attached eddy being its wall-normal height as per the attached-eddy model \citep{Townsend1976,Perry1982}, it is self-evident that it is of great importance to assess the SIAs of attached eddies with any heights in the logarithmic region, not only for the completeness of attached-eddy hypothesis, but also the accuracy of turbulence simulations \citep{Marusic2001,carper2004}. In the present study, we aim to achieve this goal by leaning upon the modified spectral stochastic estimation (SSE) proposed by \cite{Baars2016}, and dissecting the direct numerical simulations (DNS) database spanning broad-band Reynolds numbers. We will also discuss the relationship between the mean SIA and
the scale-based SIA.

\section{DNS database and methodology to calculate the SIA}
\subsection{DNS database}
The DNS database adopted in the present study has been extensively validated by Jim\'enez and co-workers \citep{DelAlamo2003, DelAlamo2004, Hoyas2006, Lozano-Duran2014}. Four cases at
$Re_\tau$=545, 934, 2003 and 4179 are used and named as Re550, Re950, Re2000 and Re4200, respectively ($Re_{\tau}=hu_{\tau}/\nu$, $h$ denotes the channel half-height, $u_{\tau}$ the wall friction velocity and $\nu$ the kinematic viscosity). 
All these data are provided by the Polytechnic University of Madrid.
Details of the parameter settings are listed in Table \ref{tab:grid}. Note that the relatively smaller computational domain size of Re4200 may influence the estimation of SIAs of the attached eddies populating the upper part of the logarithmic region. This limitation will be discussed in section \ref{re} and Appendix A.

\begin{table}
\begin{center}
	\def~{\hphantom{0}}
	\begin{tabular}{lcccccccccc}
		$Case$   & $Re_\tau$   &   $L_x(h)$&   $L_y(h)$ & $L_z(h)$ & $\Delta x^+$ & $\Delta z^+$ & $\Delta y_{min}^+$ & $\Delta y_{max}^+$ & $N_F$ & $Tu_{\tau}/h$ \\[3pt]
		Re550   & 547 & $8\pi$ &2& $4\pi$    & 13.4 & 6.8 & 0.04 & 6.7 & 142 & 22\\
		Re950   & 934 & $8\pi$ &2& $3\pi$  & 11.5 & 5.7 & 0.03 & 7.6 & 73 & 12\\
		Re2000  & 2003 & $8\pi$ &2& $3\pi$  & 12.3 & 6.2 & 0.32 & 8.9 & 48 & 11\\
		Re4200  & 4179 & $2\pi$ &2& $\pi$  & 12.3 & 6.2 & 0.32 & 10.6 & 40 & 15\\
	\end{tabular}
	\caption{Parameter settings of the DNS database. Here, $L_x$, $L_y$ and $L_z$ are the sizes of the computational domain
		in the streamwise, wall-normal and spanwise
		directions, respectively.  $\Delta x^+$ and $\Delta
		z^+$ denote the streamwise and spanwise grid
		resolutions in viscous units, respectively. $\Delta
		y_{min}^+$ and $\Delta y_{max}^+$ denote the finest and
		coarsest resolution in the wall-normal direction,
		respectively. $N_F$ and $Tu_{\tau}/h$ indicate the number of instantaneous flow fields and the total eddy turnover time used to accumulate statistics, respectively.
	}
	\label{tab:grid}
\end{center}
\end{table}

\subsection{Spectral stochastic estimation}
According to the inner-outer interaction model \citep{Marusic2010}, the large-scale motions would exert the footprints on the near-wall region, i.e., the superposition effects. \cite{Baars2016} demonstrated that this component (denoted as $u_{L}^{'+}(x^{+}, y^{+}, z^{+})$) can be obtained by the spectral stochastic estimation of the streamwise velocity fluctuation at the logarithmic region $y_o^+$, namely by 
\begin{equation}
u_{L}^{'+}\left(x^{+}, y^{+}, z^{+}\right)=F_{x}^{-1}\left\{H_{L}\left(\lambda_{x}^{+}, y^{+}\right) F_{x}\left[u_{o}^{'+}\left(x^{+}, y_{o}^{+}, z^{+}\right)\right]\right\},
\end{equation}
where $u_{o}^{'+}$ is the streamwise velocity fluctuation at $y_o^+$ in the logarithmic region, and, $F_x$ and $F_x^{-1}$ denote the FFT and the inverse FFT in the streamwise direction, respectively. $H_L$ is the transfer kernel, which evaluates the correlation between $\hat{u'}(y^+)$ and $\hat{u_o'}(y_o^+)$ at a given length scale $\lambda_{x}^{+}$, and can be calculated as
\begin{equation}\label{HL}
H_{L}\left(\lambda_{x}^{+},y_o^{+}\right)=\frac{\left\langle\hat{u'}\left(\lambda_{x}^{+}, y^{+}, z^{+}\right) \overline{\hat{u_o'}}\left(\lambda_{x}^{+}, y_{o}^{+}, z^{+}\right)\right\rangle}{\left\langle\hat{u_o'}\left(\lambda_{x}^{+}, y_{o}^{+}, z^{+}\right) \overline{\hat{u_o'}}\left(\lambda_{x}^{+}, y_{o}^{+}, z^{+}\right)\right\rangle},
\end{equation}
where $\hat{u'}$ is the Fourier
coefficient of $u'$, and $\overline{\hat{u'}}$ is the complex conjugate of $\hat{u'}$. $y^+$ is set
as $y^+=0.3$, and the outer reference height $y_o^+$ varies from $100$ to the outer region $0.7h^+$ according to the wall-normal grid distribution. Once $u_L^{'+}$ is obtained, the superposition component of $\tau_x^{'+}$ can be calculated by definition (i.e., $\frac{ \partial u_L^{'+}} { \partial y^+}$ at the wall) and denoted as $\tau_{x,L}^{'+}(y_o^+)$. 

Analogously, to eliminate the effects from the wall-detached eddies with random orientations, which contribute significantly to the streamwise velocity fluctuations at $y_o^+$, we can also use the near-wall streamwise velocity fluctuation in viscous layer $y^+$ to reconstruct the wall-coherent streamwise velocity fluctuation in the logarithmic region $y_o^+$ by spectral stochastic estimation \citep{adrian1979}, i.e.,
\begin{equation}\label{uw}
u_{W}^{'+}\left(x^{+}, y_o^{+}, z^{+}\right)=F_{x}^{-1}\left\{H_{W}\left(\lambda_{x}^{+}, y^{+}\right) F_{x}\left[u^{'+}\left(x^{+}, y^{+}, z^{+}\right)\right]\right\},
\end{equation}
where $u_{W}^{'+}$ is the wall-coherent component of $u_{o}^{'+}$.  
The wall-based transfer kernel $H_W$ can be calculated as
\begin{equation}\label{HW}
H_{W}\left(\lambda_{x}^{+},y_o^{+}\right)=\frac{\left\langle\hat{u_o'}\left(\lambda_{x}^{+}, y_o^{+}, z^{+}\right) \overline{\hat{u'}}\left(\lambda_{x}^{+}, y^{+}, z^{+}\right)\right\rangle}{\left\langle\hat{u'}\left(\lambda_{x}^{+}, y^{+}, z^{+}\right) \overline{\hat{u'}}\left(\lambda_{x}^{+}, y^{+}, z^{+}\right)\right\rangle}.
\end{equation}

\begin{figure} 
\centering 
\subfigure{ 
	\includegraphics[width=5.0in]{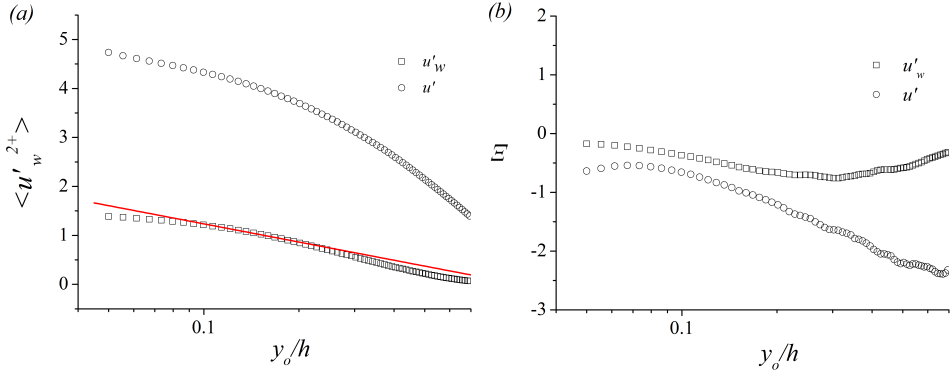} 
}
\caption{($a$) Variation of the statistic $\langle u_{W}^{'2+} \rangle$  as a function of $y_o/h$, and the full-channel data $\langle u_{}^{'2+} \rangle$ are included for comparison; ($b$) variations of the indicator functions $\Xi$ as  functions of $y_o/h$. The red line in ($a$) denotes the logarithmic decaying Eq.~(\ref{LV}) with $C_1=0.54$.}
\label{fig:uw} 
\end{figure}

Fig.~\ref{fig:uw}($a$) shows the variation of  $\langle u_{W}^{'2+} \rangle$ as a function of $y_o/h$ in the case Re2000. The full-channel data are included for comparison. It can be seen that $\langle u_{W}^{'2+} \rangle$ roughly follows the logarithmic decay for $0.09\le y_o/h\le0.2$, i.e., the logarithmic region. To quantify the logarithmic decay systematically, we define the indicator function $\Xi=y(\partial \langle u^{'2+} \rangle/\partial y)$, and display their variations in Fig.~\ref{fig:uw}($b$). Comparing with the full-channel data, a comparatively well-defined plateau is observed for $\langle u_{W}^{'2+} \rangle$. The logarithmic variance of $\langle u_{W}^{'2+} \rangle$ shown in Fig.~\ref{fig:uw}
is the consequence of the additive attached eddies \citep{Townsend1976}, and can be expressed as
\begin{equation}\label{LV}
\langle u_{W}^{'2+} \rangle=C_2-C_1\ln(y_o/h),
\end{equation}
where $C_2$ and $C_1$ are two constants, and $C_1$ is approximately equal to 0.54. Actually, the magnitude of the slope of the logarithmic decaying is affected by the Reynolds number, the configuration of the wall turbulence, the methodology for isolating the signals carried by the attached eddies, and the effects of the VLSMs. The indicator function $\Xi$ of the fully-channel data shown in Fig.~\ref{fig:uw}($b$) suggests that the logarithmic region of case Re2000 is not full-developed, as  the slope value of the logarithmic decaying is smaller than the Townsend-Perry constant $1.26$ reported at high-Reynolds number experiments \citep{Marusic2013}, and close to the magnitude of $C_1$ observed here. Furthermore, \cite{Baars2020} reported that $C_1=0.98$ in turbulent boundary layers by analyzing the streamwise velocity fluctuations carried by the attached eddies in the logarithmic region, while \cite{Hu2020} and \cite{Hwang2020} showed that $C_1=0.8$ and 0.37 in channel flows, respectively. \cite{Hu2020} adopted a scale-based filter to extract the streamwise velocity fluctuations associated with the attached eddies in the logarithmic region and did not take their imperfect coherence with the near-wall flow at each scale into account. The wall-based transfer kernel $H_W$ in Eq.~(\ref{HW}) employed here can  achieve this. \cite{Hwang2020} utilized the three-dimensional clustering method to identify the wall-attached structures in a channel flow. The differences among these decomposition methodologies may be the reason why the magnitude of $C_1$ for turbulent channel flows reported by \cite{Hu2020} and \cite{Hwang2020} is not identical to that of the present study. Besides, it is noted that the effects of VLSMs are also retained in $\langle u_{W}^{'2+} \rangle$, and their impacts on the logarithmic decaying are non-negligible. By the way, 
the methodology introduced in section \ref{Mt} to estimate the  SIAs of attached eddies at a single scale can effectively diminish the effects originating from the VLSMs (see Fig.~\ref{fig:CS}). In summary, these observations demonstrate that $ u_{W}^{'}$ can be approximately considered as the streamwise velocity fluctuations carried by the multi-scale wall-attached eddies. We will focus on the statistics in the logarithmic region in the following sections.

\subsection{Methodology to isolate targeted eddies }\label{Mt}
Apparently, the SIAs of attached eddies at a single scale ($\alpha_s$) can not be pursued by Eq.~(\ref{corr})-(\ref{mangle}). It is worth noting that in Eq.~(\ref{corr})-(\ref{mangle}), the input parameter and signals are $y_o$, $\tau_x'$ and $u'(y_o)$. Thus, to obtain an accurate $\alpha_s$, $y_o$ should be set reasonably, and $\tau_x'$ and $u'(y_o)$ should also be properly processed, to characterize the properties of the attached eddies at the targeted scale. Our new approach is based on this understanding.

According to the hierarchical distribution of the multi-scale attached eddies in high-Reynolds number wall turbulence (see Fig.~\ref{fig:AEH}($b$), also Fig. 14 of \cite{Perry1982}), $\tau_{x,L}^{'+}(y_o^+)$ represents the superposition contributed from the wall-attached motions with their height larger than $y_o^+$. Thus, the difference value $\Delta \tau_{x,L}^{'+}(y_o^+)=\tau_{x,L}^{'+}(y_o^+)-\tau_{x,L}^{'+}(y_o^++\Delta y^+)$ can be interpreted as the superposition contribution generated by the wall-attached eddies with their wall-normal heights between $y_o^+$ and $y_o^++\Delta y^+$. Here, $y_o^++\Delta y^+$ is the location of the adjacent wall-normal grid cell of that at $y_o^+$, as $\Delta y^+$ is the local grid spacing along the wall-normal direction, in viscous units, and determined by the simulation setups. The similar numerical framework has been
verified by our previous study \citep{Cheng2022}. Correspondingly, the difference value $\Delta u_{W}^{'+}(y_o^+)=u_{W}^{'+}(y_o^+)-u_{W}^{'+}(y_o^++\Delta y^+)$ is the streamwise
velocity fluctuation carried by attached eddies populating the region between $y_o^+$ and $y_o^++\Delta y^+$. In this way, the SIAs of these eddies can be assessed by
\begin{equation}\label{sangle}
\alpha_s(y_m)=\arctan(\frac{y_m}{\Delta x_{p}}),
\end{equation}
where $y_m=\frac{y_o+(y_o+\Delta y)}{2}$, and $\Delta x_{p}$ is the streamwise delay associated with the peak of the cross correlation
\begin{equation}\label{corrnew}
R_{LW}(\Delta x)=\frac{\langle\Delta \tau_{x,L}^{'+}(x,y_o^+) \Delta u_{W}^{'+}(x+\Delta x,y_o^+)\rangle}{\sqrt{\left\langle\Delta \tau_{x,L}^{'+2}\right\rangle\left\langle \Delta u_{W}^{'+2}\right\rangle}}.
\end{equation}
\begin{figure} 
\centering 
\subfigure{ 
	\includegraphics[width=2.5in]{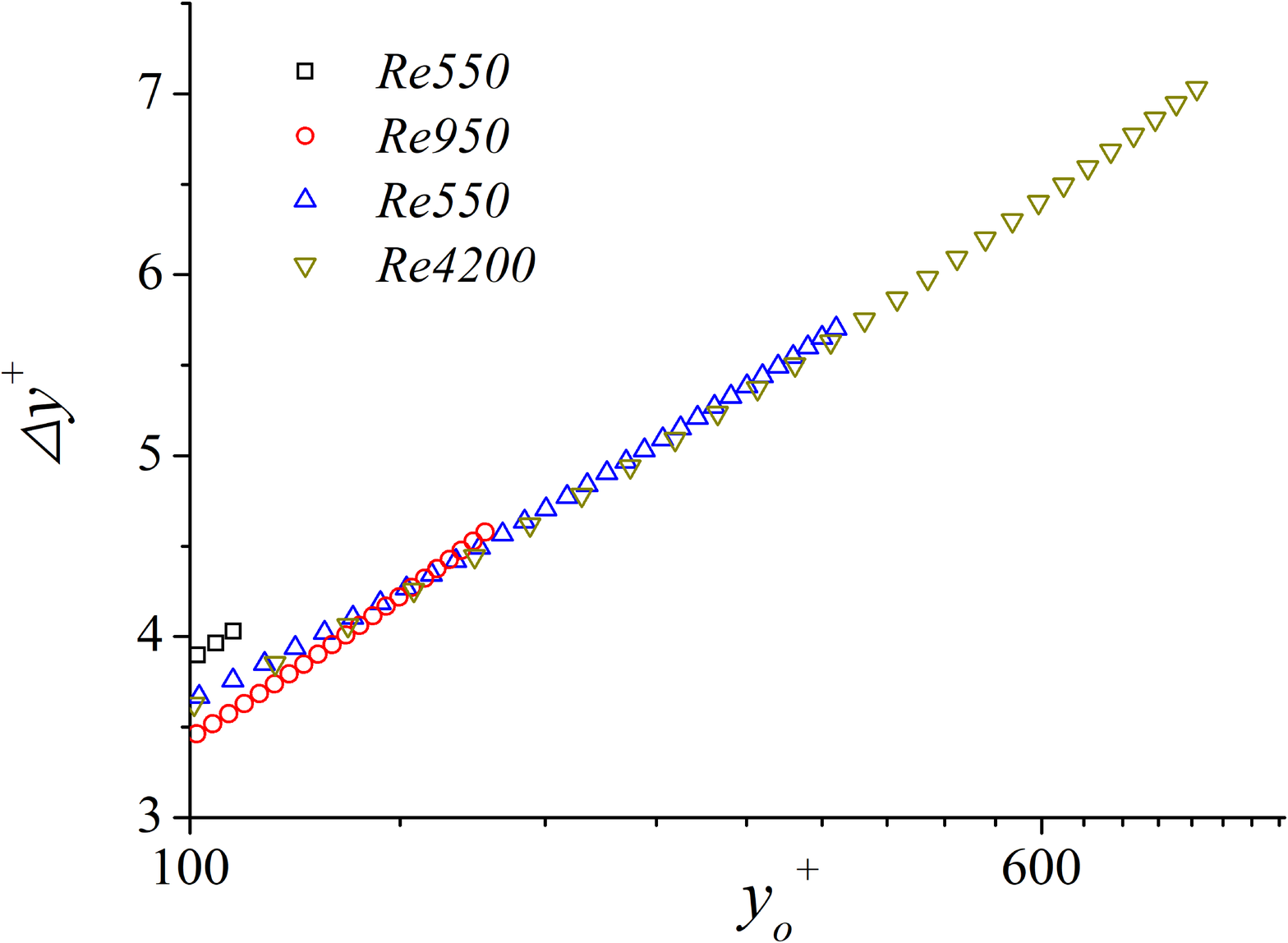} 
}
\caption{Variations of $\Delta y^+$ as functions of $y_o^+$ in the logarithmic region for all cases.}
\label{fig:dy} 
\end{figure}

As the statistical characteristics of an individual attached eddy being self-similar with its wall-normal height as per the attached-eddy hypothesis \citep{Townsend1976}, $y_m$ is just the characteristic scale of the wall-attached motions within $y_o^+$ and $y_o^++\Delta y^+$. Fig.~\ref{fig:dy} shows the variations of $\Delta y^+$ as functions of $y_o^+$ in the logarithmic region for all cases. It can be seen that the maximum values of $\Delta y^+$ are less than 7 in the case Re4200. In this regard, treating $y_m$ as the mean height of the attached eddies populating the region between $y_o$ and $y_o+\Delta y$ is reasonable, as the zone between $y_o$ and $y_o+\Delta y$ is narrow compared to the spanning of the logarithmic region. The new procedure isolates the attached eddies at a given scale from the rest of the turbulence. The cross correlation, i.e., Eq.~(\ref{corrnew}), gets rid of the influences originated from other scales, and preserves the phase information of the wall-attached motions with wall-normal height $y_m$.

At last, the critical assumptions of the present approach and its realization merit a discussion. Our methodology is based on the hierarchical distribution of the attached eddies, and the hypothesis that the characteristic velocity scales carried by the attached eddies with different wall-normal heights are identical with their scale interactions omitted. That's to say, the attached eddies in each hierarchy contribute equally to the streamwise wall-shear fluctuations on the wall surface and the streamwise turbulence intensity in the lower bound of the logarithmic region. Only in this way, both $\Delta \tau_{x,L}^{'+}$ and $\Delta u_{W}^{'+}$ 
approximately reflect the characteristics of the attached eddies at $y_m$. 
In fact, these assumptions are also the key elements when developing the attached-eddy model \citep{Townsend1976,Perry1982,Woodcock2015,Yang2016a,Mouri2017,Yang2017}, and some of them may be valid only in high-Reynolds number wall turbulence. For example, the hierarchical distribution of the multi-scale attached eddies is prominent at high-Reynolds number turbulence \citep{DeSilva2016,Marusic2019,Cheng2019}.
However, when the DNS data listed in Table \ref{tab:grid} are utilized to study the characteristics of the attached eddies, the finite Reynolds-number effects and the intricate scale interactions would take effects inevitably. Besides, the VLSMs, which can not be depicted by the attached-eddy model, would also impose non-trivial impacts \citep{Perry1995,Hwang2020,Baars2020a}. Accordingly, the subtraction between $u_{W}^{'+}(y_o^+)$ and $u_{W}^{'+}(y_o^++\Delta y^+)$ can not achieve a sharp cut-off at the targeted scale in the spectral space, and hereby the spectrum of $\Delta u_{W}^{'+}(y_o^+)$ would be comparatively small but not negligible at the smaller and larger scales of the targeted one. 
The finiteness of $\Delta y^+$ is another factor, which is worth attention in some scenarios. Due to the limitations of numerical simulation, $\Delta y^+$ is a finitely small quantity. When assessing the SIA of the attached eddies at a given wall-normal height, treating $y_m^+$ as their characteristic scales (therefore, neglecting the effects of the narrowband between $y^+$ and $y^++\Delta y^+$) is acceptable, because $\Delta y^+$ is rather small compared to the spanning of the whole logarithmic region. The linear growth of the typical length scales of $\Delta \tau_{x,L}^{'+}$ and $\Delta u_{W}^{'+}$ shown in Fig.~\ref{fig:Ruu}($b$) can verify this validity. On the other hand, when the spectral characteristics of $\Delta u_{W}^{'+}$ are considered, $\Delta u_{W}^{'+}$ should be interpreted as the additive outcomes of the attached eddies with their wall-normal heights within $y^+$ and $y^++\Delta y^+$, strictly speaking. Under this circumstance, the spectral energy distribution that corresponds to the self-similar attached eddies within this range should be observed to peak around the dominate wavelength and vary continuously and locally. The results shown in Fig.~\ref{fig:spk1} confirm our proposition. Details will be discussed in the following section.

\section{Results}\label{re}
\begin{figure} 
\centering 
\subfigure{ 
	\label{fig:CS:a} 
	\includegraphics[width=5.0in]{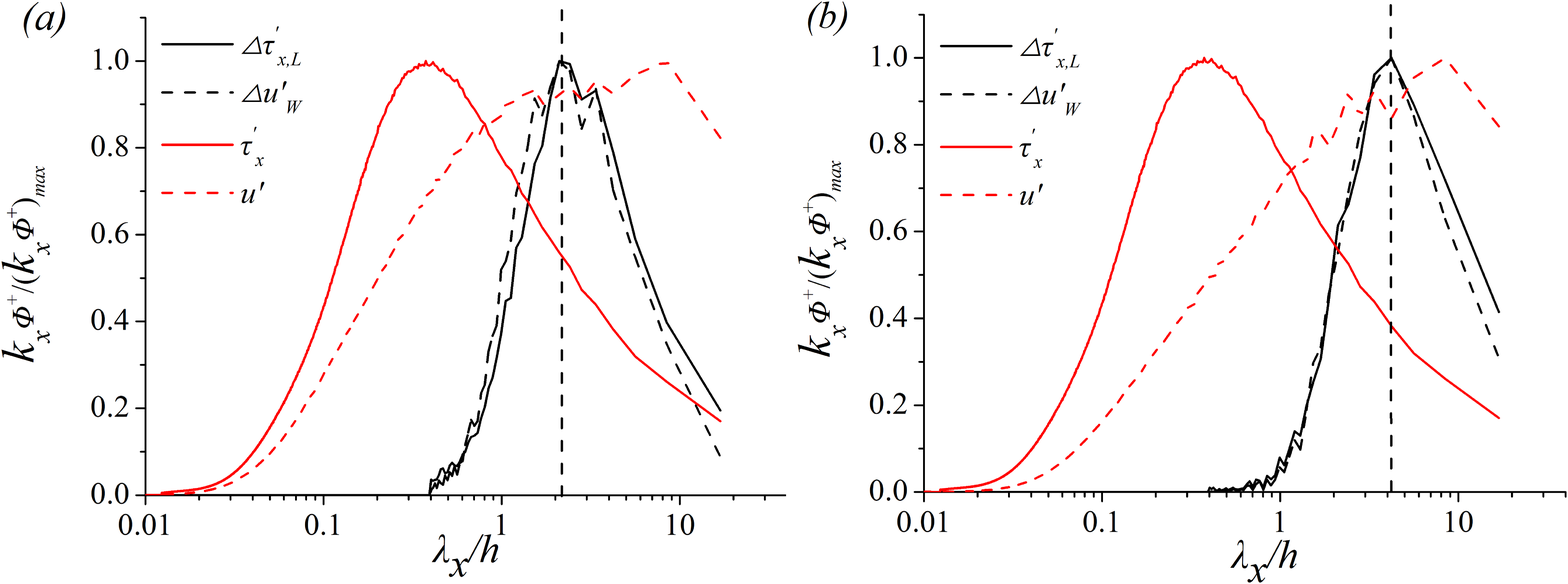} 
}
\caption{Streamwise premultiplied spectra of $\Delta \tau_{x,L}^{'}$ , $\Delta u_{W}^{'}$, $\tau_{x}^{'}$ and $u'$ for ($a$) $y_o=0.1h$ and
	($b$) $y_o=0.2h$ in the case Re2000. Each spectrum is normalized with its maximum value. The vertical dashed lines in ($a$) and ($b$) are plotted to highlight the corresponding $\lambda_x/h$ of the maximum values of the premultiplied spectra of $\Delta \tau_{x,L}^{'}$ and $\Delta u_{W}^{'}$.}
\label{fig:CS} 
\end{figure}

Before investigating the SIAs of attached eddies, it is important to study the characteristic scales of $\Delta \tau_{x,L}^{'}$ and $\Delta u_{W}^{'}$ first. Figs.~\ref{fig:CS}($a$) and \ref{fig:CS}($b$) show their streamwise premultiplied spectra at $y_o=0.1h$ and $y_o=0.2h$ for Re2000, respectively. The spectra of $\tau_{x}^{'}$ and $u'$ of the full-channel data are also included for comparison. Each spectrum is normalized with its maximum value. It can be seen that the spectra of $\Delta \tau_{x,L}^{'}$ and $\Delta u_{W}^{'}$ are roughly coincident, and peak at $\lambda_x=2.1h$ for $y=0.1h$, and $\lambda_x=4.2h$ for $y=0.2h$, respectively. By contrast, the spectra of $\tau_{x}^{'}$ and $u'$ do not share similar spectral characteristics. It is noted that $\Delta u_{W}^{'2}$ and $\Delta \tau_{x,L}^{'2}$ only account for very little energy of the full-channel signals at the same wall-normal positions. For example, $\Delta u_{W}^{'2}$ at $y_o=0.1h$ and $y_o=0.2h$ occupies 0.0034$\%$ and 0.002$\%$ of $u^{'2}$ at the corresponding positions, respectively, whereas $\Delta \tau_{x,L}^{'2}$ for $y_o=0.1h$ and $y_o=0.2h$ occupies 0.012$\%$ and 0.0045$\%$ of $\tau_{x}^{'2}$, respectively. Moreover, comparing with the spectra of the full-channel data, the spectra of $\Delta u_{W}^{'}$ decay rapidly when $\lambda_x\geq4h$ (see Figs.~\ref{fig:CS}), which indicates that the effects of VLSMs on $\Delta u_{W}^{'}$  are rather 
limited.

\begin{figure} 
\centering 
\subfigure{ 
	\includegraphics[width=5.0in]{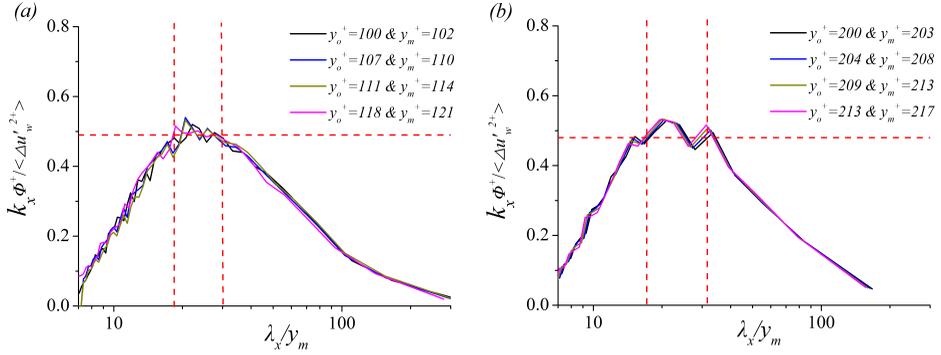} 
}
\caption{Premultiplied one-dimensional streamwise spectra of $\Delta u_{W}^{'}$ around ($a$) $y_o=0.05h$; ($b$) $y_o=0.1h$ in Re2000. The horizontal dashed lines represent the plateaus or peaks of the spectra. The vertical lines are plotted to highlight the self-similar regions of each spectrum.}
\label{fig:spk1} 
\end{figure}

Fig.~\ref{fig:spk1} shows the streamwise premultiplied spectra of $\Delta u_{W}^{'}$ around $y_o=0.05h$ and $y_o=0.1h$. Each spectrum is normalized by the energy of $\Delta u_{W}^{'}$ at a given $y_m$. Clear plateau regions can be observed around the spectral peaks. For $y_o=0.05h$, the region is between $18\le\lambda_x/y_m \le30$, and for $y_o=0.1h$, it is between $17\le\lambda_x/y_m \le31$ , which corresponds to the $k_x^{-1}$ region in the spectrum predicated by the attached-eddy model, and can be considered as the spectral signatures of the attached eddies \citep{Perry1982,Perry1986,Hwang2020,Deshpande2021a}. Besides, the spectra shown here resemble the spectrum of the type
A eddies hypothesized by \cite{Marusic1995}, i.e., the energy fraction captured by the attached-eddy model. These observations support the proposition that the $\Delta u_{W}^{'}$ signals are the streamwise velocity fluctuations carried by the self-similar attached eddies predominantly. Moreover, they also indicate that the streamwise length scales of the dominant eddies increase with $y_o$, as the self-similar range is not altered significantly with increasing $y_o$.

\begin{figure} 
\centering 
\subfigure{ 
	\label{fig:Ruu:a} 
	\includegraphics[width=5.0in]{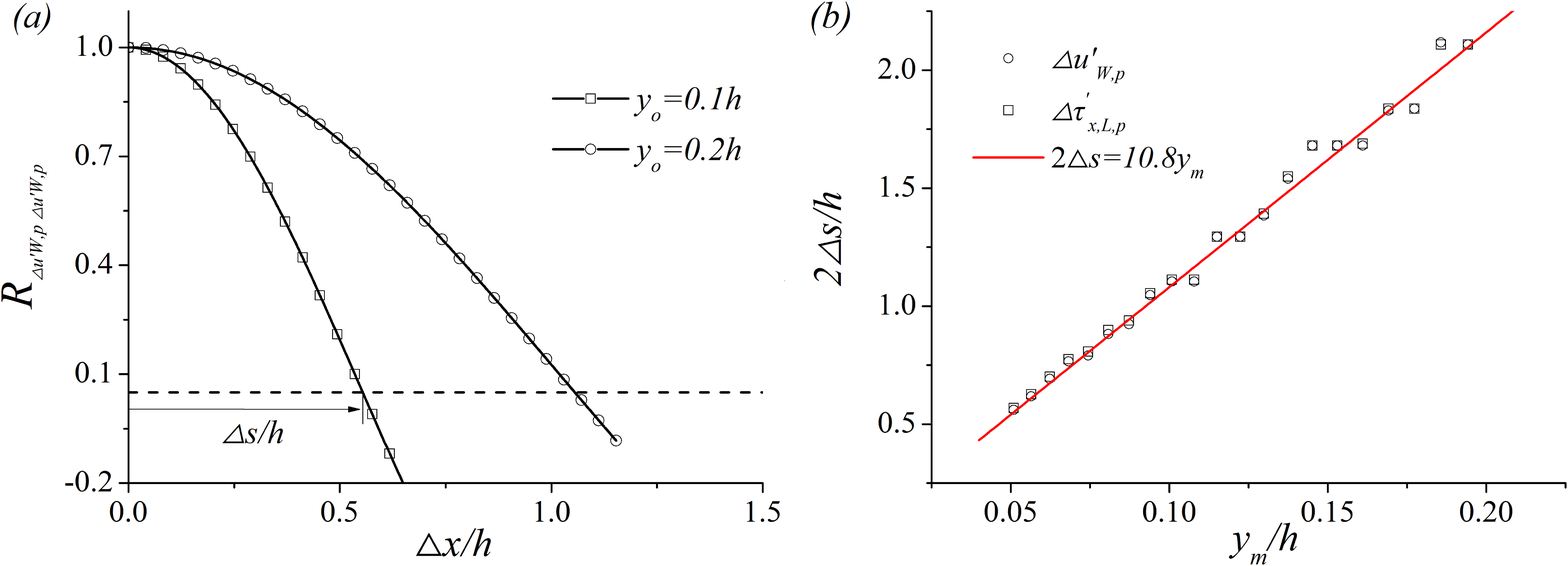} 
}
\caption{($a$) Variations of $R_{\Delta u_{W,p}^{'}\Delta u_{W,p}^{'}}$ as functions of $\Delta x/h$ for two selected $y_o$; ($b$) variations of $\Delta s/h$ as functions of $y_m^+$ for $\Delta \tau_{x,L,p}^{'+}$ and $\Delta u_{W,p}^{'+}$. The line in ($b$) denotes the linear variation $2\Delta s=10.8y_m$.}
\label{fig:Ruu} 
\end{figure}

To further investigate the scale characteristics of $\Delta \tau_{x,L}^{'}$ and $\Delta u_{W}^{'}$, the autocorrelation function of $\Delta u_{W,p}^{'+}$ (the signals that are extracted from the spectral peaks shown in Fig.~\ref{fig:spk1}, i.e., filtered $\Delta u_{W}^{'}$ with wavelength larger than $17y_m$, but smaller than $31y_m$) is considered,  which takes the form of
\begin{equation}
R_{\Delta u_{W,p}^{'}\Delta u_{W,p}^{'}}(\Delta x,y_o)=\frac{\langle\Delta u_{W,p}^{\prime}\left(x, y_o, z\right)\Delta u_{W,p}^{\prime}\left(x+\Delta x, y_o, z\right)\rangle }{\langle\Delta u_{W,p}^{\prime 2}\left(x, y_o, z\right)\rangle},
\end{equation}
and the counterpart of $\Delta \tau_{x,L}^{'+}$ can also be defined similarly. Fig.~\ref{fig:Ruu}($a$) shows the variations of $R_{\Delta u_{W,p}^{'}\Delta u_{W,p}^{'}}$ as functions of $\Delta x/h$ for two selected $y_o$. The larger $y_o$, the broader the $R_{\Delta u_{W,p}^{'}\Delta u_{W,p}^{'}}$. As a measure of the typical length scale, we employ $\Delta s/h$, which is the streamwise delay corresponding to $R_{\Delta u_{W,p}^{'}\Delta u_{W,p}^{'}}=0.05$ or $R_{\Delta \tau_{x,L,p}^{'}\Delta \tau_{x,L,p}^{'}}=0.05$ (here, 0.05 is an empirical small positive threshold). 
Fig.~\ref{fig:Ruu}($b$) shows the variations of $2\Delta s/h$ as functions of $y_m/h$ for $\Delta \tau_{x,L,p}^{'+}$ and $\Delta u_{W,p}^{'+}$. 
For both $\Delta \tau_{x,L,p}^{'+}$ and $\Delta u_{W,p}^{'+}$, $2\Delta s/h$ increases linearly with $y_m/h$ throughout most of the logarithmic region. This observation is consistent with the attached-eddy hypothesis, which states that the length scales of the attached eddies grow linearly with their wall-normal heights \citep{Hwang2015,Marusic2019}. Moreover, both the streamwise length scales of $\Delta \tau_{x,L,p}^{'+}$ and $\Delta u_{W,p}^{'+}$ follow $2\Delta s=10.8y_m$ (consider the symmetry of the autocorrelation function with respect to $\Delta x=0$, $2\Delta s$ truly represents the streamwise length scale of the signals). This scale characteristic
agrees well with some previous studies. For example, \cite{Baars2017} showed that the streamwise/wall-normal aspect ratio of the wall-attached eddy structure is $\lambda_x/y=14$ in turbulent boundary layers, which is close to the result here. \cite{Hwang2020} reported that the spectra of the self-similar wall-attached structures agree with the attached-eddy hypothesis at $\lambda_x=12y$, which is consistent with the estimation of the present study. All these observations indicate that $\Delta \tau_{x,L}^{'+}$ and $\Delta u_{W}^{'+}$ are representative of  the attached eddies at a certain wall-normal height, though the minor influences of VLSMs still exist, and treating $y_m^+$ as their characteristic scales is reasonable.

In summary, all the observations mentioned above indicate that $\Delta \tau_{x,L}^{'}$ and $\Delta u_{W}^{'}$ are the outcomes of the energy-containing motions with the wall-normal heights approximately equal to $y_m$, and the cross correlation, i.e., Eq.~(\ref{corrnew}), truly reflects the phase difference between the streamwise velocity fluctuations carried by these motions and their footprints in the near-wall region. Other wall-normal positions and DNS cases yield similar results and are not shown here for brevity.

\begin{figure} 
\centering 
\subfigure{ 
	\label{fig:afa:a} 
	\includegraphics[width=5.0in]{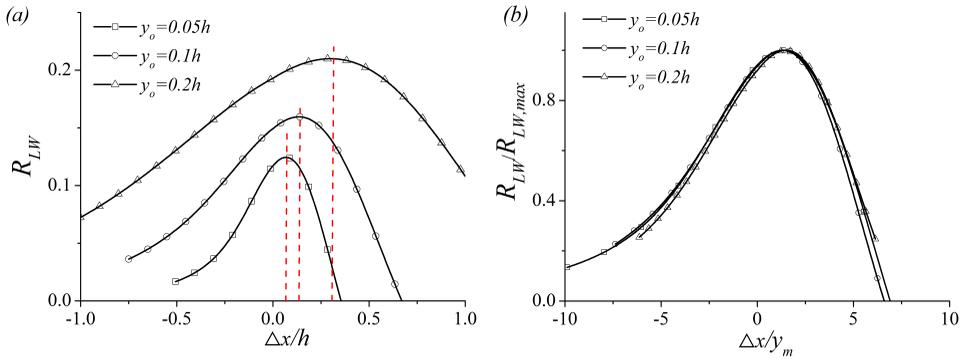} 
}
\caption{($a$) Variations of $R_{LW}$, i.e., the cross correlation between $\Delta \tau_{x,L}^{'+}(y_o^+)$ and $\Delta u_{W}^{'+}(y_o^+)$, as functions of $\Delta x$ for some selected $y_o$ in the case Re2000; ($b$) variations of the normalized $R_{LW}$ as functions of $\Delta x/y_m$ for some selected $y_o$ in the case Re2000. The $R_{LW}$ profiles are normalized with their maximum values in ($b$). The vertical dashed lines in ($a$) are plotted to highlight the maximum values of $R_{LW}$ and their corresponding $\Delta_x/h$.}
\label{fig:afa} 
\end{figure}

Fig.~\ref{fig:afa}($a$) shows the variations of $R_{LW}$ as functions of the streamwise delay for some selected wall-normal positions in the case Re2000. Since the streamwise length scales of the energy-containing motions are increased with their normal heights (see Fig.~\ref{fig:CS}), $R_{LW}$ becomes wider about the peak with increasing $y_o$. $\Delta x_{p}$ can be identified obviously from the cross-correlation profiles, and the SIAs of the attached eddies at a given wall-normal
height can be calculated according to Eq.~(\ref{sangle}). Fig.~\ref{fig:afa}($b$) plots the variations of the normalized $R_{LW}$ as functions of $\Delta x/y_m$ for some selected $y_o$ in the case Re2000. The $R_{LW}$ distributions are normalized with their maximum values $R_{LW,max}$. It can be seen that the profiles of $R_{LW}/R_{LW,max}$ for different wall-normal heights coincide well with each other, which indicates the self-similar characteristics of the energy-containing motions in the logarithmic region. We have checked that the correlations calculated from the raw data, i.e., $R_{\tau_x' u'}$ in Eq.~(\ref{corr}) can not coincide if normalized in this manner.
Again, it demonstrates that the new methodology is capable of capturing the main properties of the attached eddies.

\begin{figure} 
\centering 
\subfigure{ 
	\includegraphics[width=2.5in]{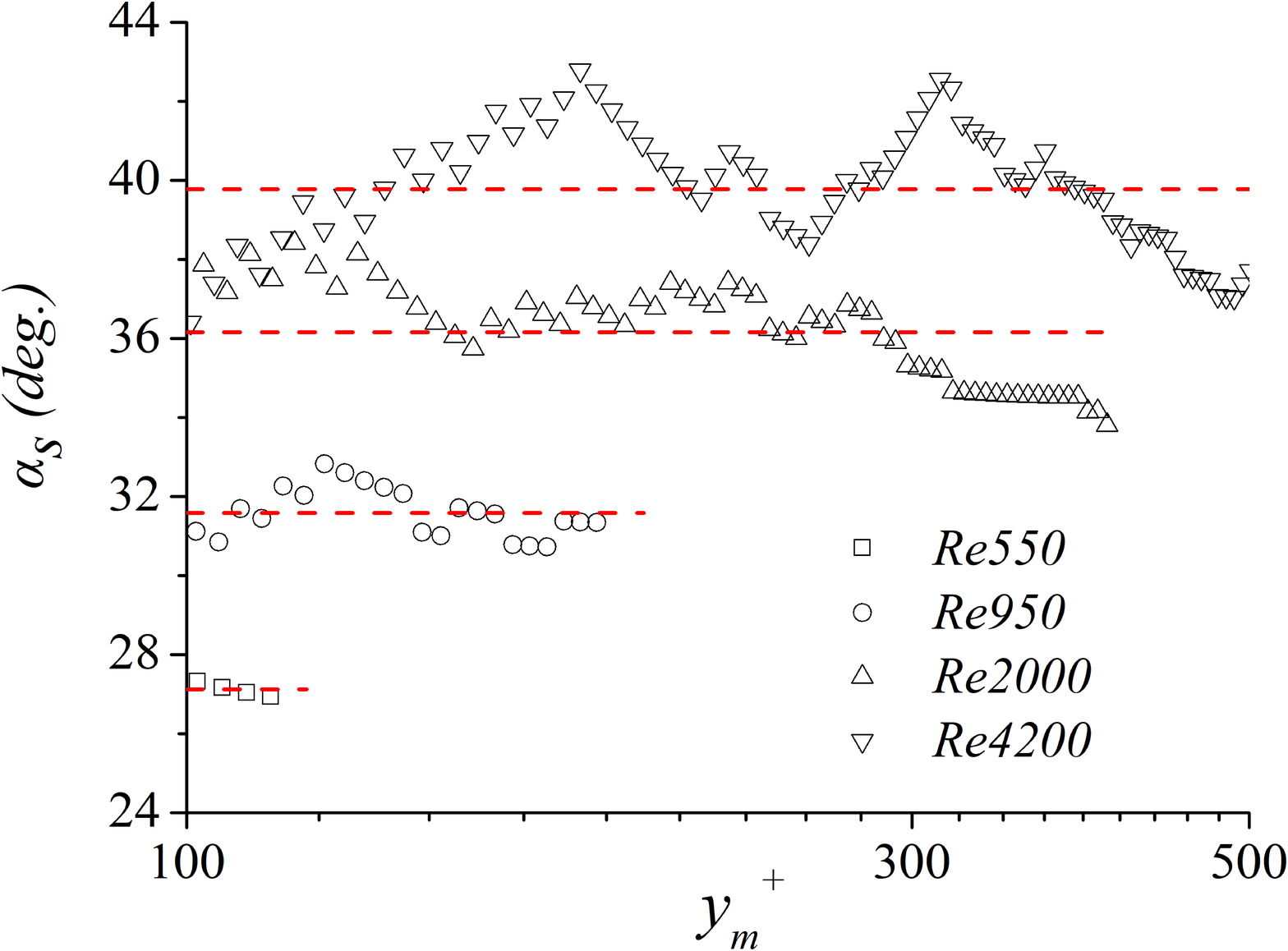} 
}
\caption{$\alpha_s$ as functions of $y_m^+$ for all cases, and the red dashed lines denote the mean $\alpha_s$ across the logarithmic region of each case.}
\label{fig:afa2} 
\end{figure}

Fig.~\ref{fig:afa2} plots the variations of $\alpha_s$ as functions of $y_m^+$ for all cases. $\alpha_s$ increases approximately from $27^{\circ}$ for Re550, to $40^{\circ}$ for Re4200.
For a given case, $\alpha_s$ changes little spanning the logarithmic region except for the upper part of logarithmic region in Re4200.
\cite{Deshpande2019} isolated the large wall-attached structures in a DNS of turbulent boundary layer at $Re_{\tau}\approx2000$, and found the corresponding SIAs to be $32^{\circ}$ (see Fig. 4($a$) of their paper).
Their observation is consistent with the results of the present study. However, \cite{Deshpande2019} only calculated the SIAs of the largest wall-attached motions in the logarithmic region due to the limitation of the methodology adopted in their study, whereas we make a thorough investigation on the SIAs of attached eddies with any wall-normal heights in the logarithmic region. Moreover, \cite{Deshpande2019} reported that the SIAs of the large wall-attached motions identified in a wind-tunnel boundary layer with $Re_{\tau}=14000$ are approximately $50^{\circ}$. They ascribed the result difference between DNS and experiment to the limited streamwise scale range owing to the DNS domain size selected for analysis. Our results reveal that the Reynolds number effects play a non-negligible role in the formation of SIAs of attached eddies. To the authors' knowledge, this is the first time that the Reynolds-number dependence of SIAs of the wall-attached motions at a given length scale has been clearly shown. At last, it should be noted that $\alpha_s$ of Re4200 decreases rapidly for $y_m^+>500$ (not shown here). This diversity is due to the small computational domain size along the streamwise direction in this database. Thus, in the discussion below,  the statistics of $\alpha_s$ in the range of $y_m^+>500$ in Re4200 will not be taken into account. The sensitivity of the presented results to the  number of instantaneous flow fields employed for accumulating statistics is examined in the Appendix A.

\begin{figure} 
\centering 
\subfigure{ 
	\label{fig:afam:a} 
	\includegraphics[width=5.0in]{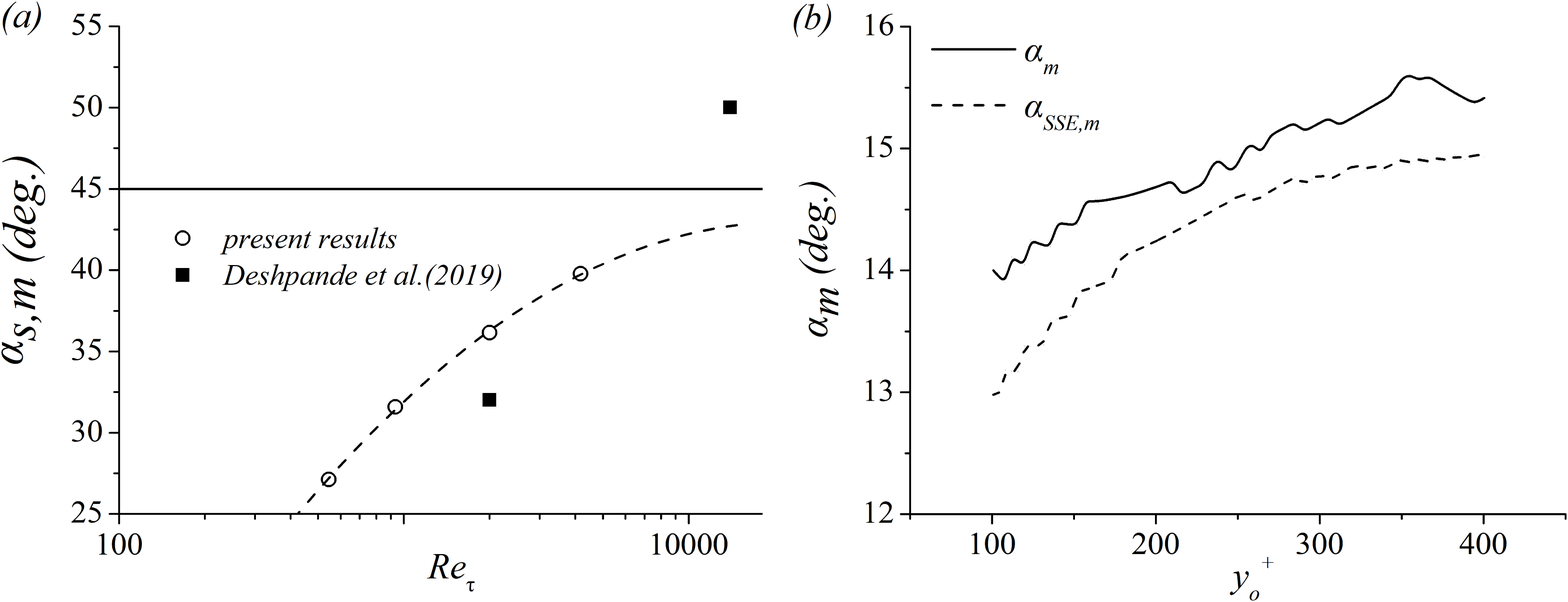} 
}
\caption{($a$) Variations of the mean $\alpha_s$ ($\alpha_{s,m}$) statistic in the range of logarithmic region as a function of the friction Reynolds number, and the experimental results of turbulent boundary layers \citep{Deshpande2019} are also included for comparison; ($b$) $\alpha_{m}$ and $\alpha_{SSE,m}$ as a function of $y_o^+$ for Re2000. The solid black line in ($a$) denotes the theoretical prediction angle $45^{\circ}$, and the dashed lines in ($a$) indicates the asymptotic behavior of $\alpha_{s,m}$.}
\label{fig:afam} 
\end{figure}

Fig.~\ref{fig:afam}($a$) shows the mean $\alpha_s$ ($\alpha_{s,m}$) distribution in the range of logarithmic region as a function of the friction Reynolds number. It can be seen that the SIA may reach the theoretical prediction angle $45^{\circ}$ \citep{Perry1992} when $Re_{\tau} \sim O(10^4)$. The results of DNS of a turbulent boundary layer and wind-tunnel experiment of \cite{Deshpande2019} are roughly agreed with the tendency. The minor differences may result from the distinct configurations of the wall-bounded turbulence.

\section{Disscussion}
\subsection{Effects of near-wall and detached motions}
To clarify the effects of near-wall and detached motions on the SIA assessment, we
calculate the mean SIA based on the predictive signals, i.e.,
\begin{equation}\label{Imangle}
\alpha_{SSE,m}=\arctan(\frac{y_o}{\Delta x_{p}}),
\end{equation}
where $\Delta x_{p}$ is the streamwise delay associated with the peak of the following cross correlation
\begin{equation}\label{Icorr}
R_{\tau_{x,L}' u_W'}(\Delta x)=\frac{\langle\tau_{x,L}^{'}(x) u_W'(x+\Delta x,y_o)\rangle}{\sqrt{\left\langle\tau_{x,L}^{'2}\right\rangle\left\langle u_W^{'2}\right\rangle}}.
\end{equation}

Fig.~\ref{fig:afam}($b$) shows the variations of $\alpha_{SSE,m}$ as a function of $y_o^+$ for Re2000, and the statistics of $\alpha_{m}$ are also included
for comparison. We can see that $\alpha_{SSE,m}$ distribution is very closed to that of $\alpha_{m}$. It highlights the fact that the phase information
embedded in the raw signals $u'(y_o^+)$ and $\tau_{x}'$ is preserved by SSE. It also suggests that the near-wall and wall-detached motions, which
can not be captured by SSE, have a negligible impact on the magnitudes of SIA.

\subsection{$\alpha_s$ versus $\alpha_m$}
Reviewing the approach to obtain the $\alpha_m$ (i.e., Eq.~(\ref{corr})-(\ref{mangle})), the proposition that $\alpha_m$ being the mean SIA
of attached eddies manifests in three aspects: ($1$) the generation of $\tau_{x}'$ is not only the outcome of the near-wall motions, but also 
the footprints of all the wall-attached eddies \citep{Cho2018,Cheng2020a}; ($2$) $u'$ in logarithmic region results from a sum of random contributions from the wall-attached eddies with distinct characteristic length scales \citep{Yang2016a}, and a portion of contributions from the wall-detached eddies \citep{Baars2020}; ($3$) $y_o$ is
a wall-normal position located in the logarithmic region and chosen arbitrarily. As mentioned above, an array of wall-attached eddies with distinct wall-normal heights can simultaneously convect past this reference position.

\begin{figure} 
\centering 
\subfigure{ 
	\label{fig:add:a} 
	\includegraphics[width=5.0in]{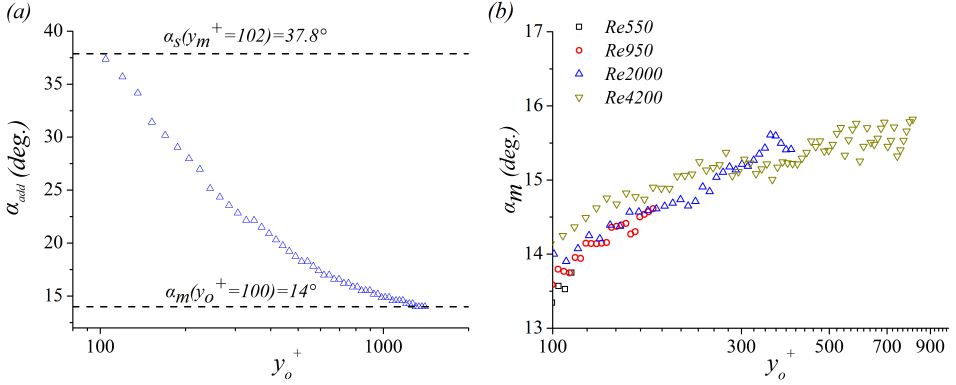} 
}
\caption{($a$) Variations of the additive SIA $\alpha_{add}$ as a function of $y_o^+$ for Re2000; ($b$) the mean SIAs $\alpha_m$ as  functions of $y_o^+$ for all cases.}
\label{fig:add} 
\end{figure}

Here, an additive SIA is calculated to highlight the relationship between $\alpha_s$ and $\alpha_m$, namely,
\begin{equation}\label{add}
\alpha_{add}=\arctan(\frac{y_s}{\Delta x_{p}}),
\end{equation}
where $y_s^+=100$ is the lower boundary of logarithmic region, and $\Delta x_{p}$ is the streamwise delay associated with the peak of the following cross correlation, i.e.,
\begin{equation}\label{corradd}
R_{add}(\Delta x)=\frac{\langle(\tau_{x,L}^{'+}(x,y_s^+)-\tau_{x,L}^{'+}(x,y_o^+)) (u_{W}^{'+}(x+\Delta x,y_s^+)-u_{W}^{'+}(x+\Delta x,y_o^+))\rangle}{\sqrt{\left\langle(\tau_{x,L}^{'+}(x,y_s^+)-\tau_{x,L}^{'+}(x,y_o^+))^{2}\right\rangle\left\langle (u_{W}^{'+}(x ,y_s^+)-u_{W}^{'+}(x,y_o^+))^{2}\right\rangle}},
\end{equation}
where the reference position $y_o^+$ varies from $y_s^++\Delta y^+$ (equals to 104) to $0.7h^+$. Fig.~\ref{fig:add}($a$) shows the variations of $\alpha_{add}$ as a function of $y_o^+$ for Re2000. It can be seen that $\alpha_{add}$ decreases from $37.8^{\circ}$ to $14^{\circ}$ as $y_o^+$ increases, which corresponds to $\alpha_{s} (y_m^+=102)$ and $\alpha_{m} (y_o^+=100)$, respectively. In other words, $\alpha_{add}$ converges from the SIAs of attached eddies with wall-normal height approximately $100$ in viscous units to the mean SIA at $y_o^+=100$. This observation can be explained through the prism of the hierarchical attached eddies in high-Reynolds number wall turbulence. The increase of $y_o^+$ indicates that $\tau_{x,L}^{'+}(y_s^+)-\tau_{x,L}^{'+}(y_o^+)$ and $u_{W}^{'+}(y_s^+)-u_{W}^{'+}(y_o^+)$ are contributed by more and more wall-attached eddies with their normal heights larger than $y_s^+$, and gradually become equal
to $\tau_{x,L}^{'+}(y_s^+)$ and $u_{W}^{'+}(y_s^+)$, respectively, when $y_o^+$ approaches $h^+$. Thus, $R_{add}$ would also gradually converge to $R_{\tau_{x,L}' u_W'}$ in Eq.~(\ref{Icorr}), and $\alpha_{add}$ converges to $\alpha_{m}$ and $\alpha_{SSE,m}$ concurrently. 

Additionally, this study helps to understand the variation tendency of  $\alpha_{m}$.  Fig.~\ref{fig:add}($b$) plots the variations  of  $\alpha_{m}$ for all cases. It is clearly observed that $\alpha_{m}$ increases continuously with $y_o^+$. Taking Re2000 as an example, $\alpha_{m}$ increases from $14^{\circ}$ for $y_s^+$ to $15.3^{\circ}$ for $y_e^+$. Increasing $y_o^+$ implies that fewer and fewer wall-attached eddies contribute to $u'$. In this way, $\alpha_{m}$ would converge to $\alpha_{s}$ as $y_o^+$ increases, albeit more slowly.

\subsection{Scale-dependent inclination angles of wall-attached eddies}
An alternative approach for calculating the scale-dependent inclination angle (SDIA) has been reported
by \cite{Baars2016}. The following are the primary processes and outcomes. The scale-specific phase between $u'$ at $y^+$ and $y_o^+$ can be estimated as
\begin{equation}
\Phi\left( \lambda_{x}\right)=\arctan\left\{\frac{\operatorname{Im}\left[\phi_{u_{o}' u_{}'}\left(\lambda_{x},y^+,y_o^+\right)\right]}{\operatorname{Re}\left[\phi_{u_{o}' u_{}'}\left(\lambda_{x},y^+,y_o^+\right)\right]}\right\},
\end{equation}
where Im($\cdot$) and Re($\cdot$) denote the imaginary and real parts of $\phi_{u_{o}' u_{}'}$, namely, the numerator of Eq.~(\ref{HL}). The scale-dependent streamwise shift can be calculated as
\begin{equation}
l(\lambda_{x})=\frac{\Phi\left( \lambda_{x}\right)\lambda_{x}}{2\pi}.
\end{equation}
Accordingly, the SDIA can be estimated as
\begin{equation}
\alpha_{sd}(\lambda_{x})=\arctan(\frac{y_o-y}{l(\lambda_{x})}).
\end{equation}
A positive $\alpha_{sd}$ value corresponds to a spatially forward-leaning structure.

\begin{figure} 
\centering 
\subfigure{ 
	\includegraphics[width=2.5in]{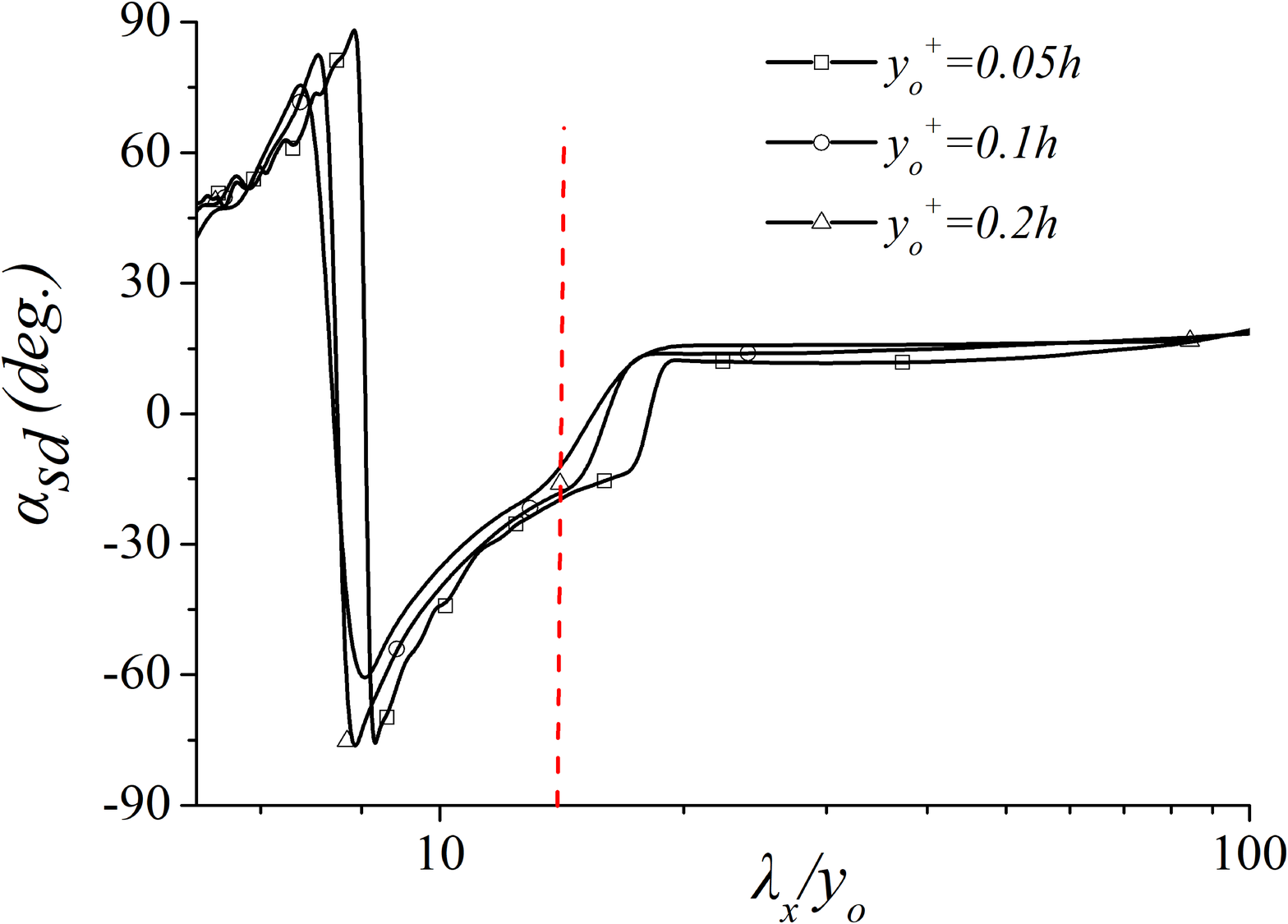} 
}
\caption{Variations of the scale-dependent inclination angles for three selected wall-normal positions in the case Re2000. The vertical line denotes $\lambda_{x}/y_o=14$.}
\label{fig:spafa} 
\end{figure}

Fig.~\ref{fig:spafa} shows the SDIAs as functions of $\lambda_{x}/y_o$ for three selected wall-normal positions in the case Re2000. For $\lambda_{x}/y_o>18$, the SDIAs of the large-scale motions are shown to be approximately equal to $14^{\circ}$ (in fact, this is not the real SIA of the large-scale wall-attached structures, according to the study of \cite{Deshpande2019}). However, for the smaller length scales, the SDIAs tend to be negative and vary rapidly with $\lambda_{x}/y_o$. This is the range of self-similar structures reported by previous studies, especially those with $\lambda_x/y_o=14$ \citep{Baars2017,Baidya2019}. Similar results have also been reported by \cite{Baars2016} (see Fig. 5 of their paper). It indicates that the  phase spectrum shown in Fig.~\ref{fig:spafa} cannot be interpreted with any physical relevance at these scales, as the scale-specific phases of them are random indeed.
The contamination from the detached eddies with random orientations could be the source of this problem. This is the main purpose of the present study, i.e., to eliminate the corruption caused by the wall-detached motions and appropriately measure the SIAs of the wall-attached eddies at a certain wall-normal height.

\section{Concluding remarks}
In the present study, we develop a methodology to assess the streamwise inclination angles of the wall-attached eddies at a given wall-normal height in turbulent channel flows, by coupling the spectral stochastic estimation with the attached-eddy hypothesis. Our results show, for the first time, that the SIAs of the attached eddies are Reynolds-number dependent in low and medium Reynolds numbers and tend to be consistent with the theoretical prediction (i.e., $\alpha_{s}=45^{\circ}$) as Reynolds number increased. We further reveal that the mean SIA reported by vast previous studies are the outcomes of the additive effect contributed by multi-scale attached eddies.

The attached-eddy model has been the guidance for the reconstruction of the velocity field in wall turbulence \citep{Perry1995,Chandran2017,Baidya2017}. Hierarchical vortex packets which consist of $\Lambda-$vortexes with $\alpha_{s}=45^{\circ}$ are distributed on the wall surface to mimic the attached eddies. The present results suggest that a lower SIA of representative structures might be helpful for a more accurate reconstruction when the Reynolds number
is not high enough. Moreover, within the state-of-the-art wall-modelled large-eddy simulation (WMLES) framework, one may estimate the instantaneous
$\tau_{x}$ based on the velocities carried by the log-region eddies \citep{Fu2021,fu2022prediction}. The Reynolds-number dependence of SIAs  of these eddies should be accounted for by an advanced model in this sense. 


\section*{Acknowledgments}
L.F. acknowledges the fund from CORE as a joint research center for ocean research between QNLM and HKUST, and the fund from Guangdong Basic and Applied Basic Research Foundation (No. 2022A1515011779).
We would like to thank Professor Jim\'enez for making the DNS data available. We also express our gratitude to the reviewers of this paper for their kind and constructive comments.

\section*{Declaration of interests}
The authors report no conflict of interest.

\section*{Appendix A. Statistic sensitivity to $N_F$}\label{Ap}
\begin{figure} 
\centering 
\subfigure{ 
	\includegraphics[width=2.5in]{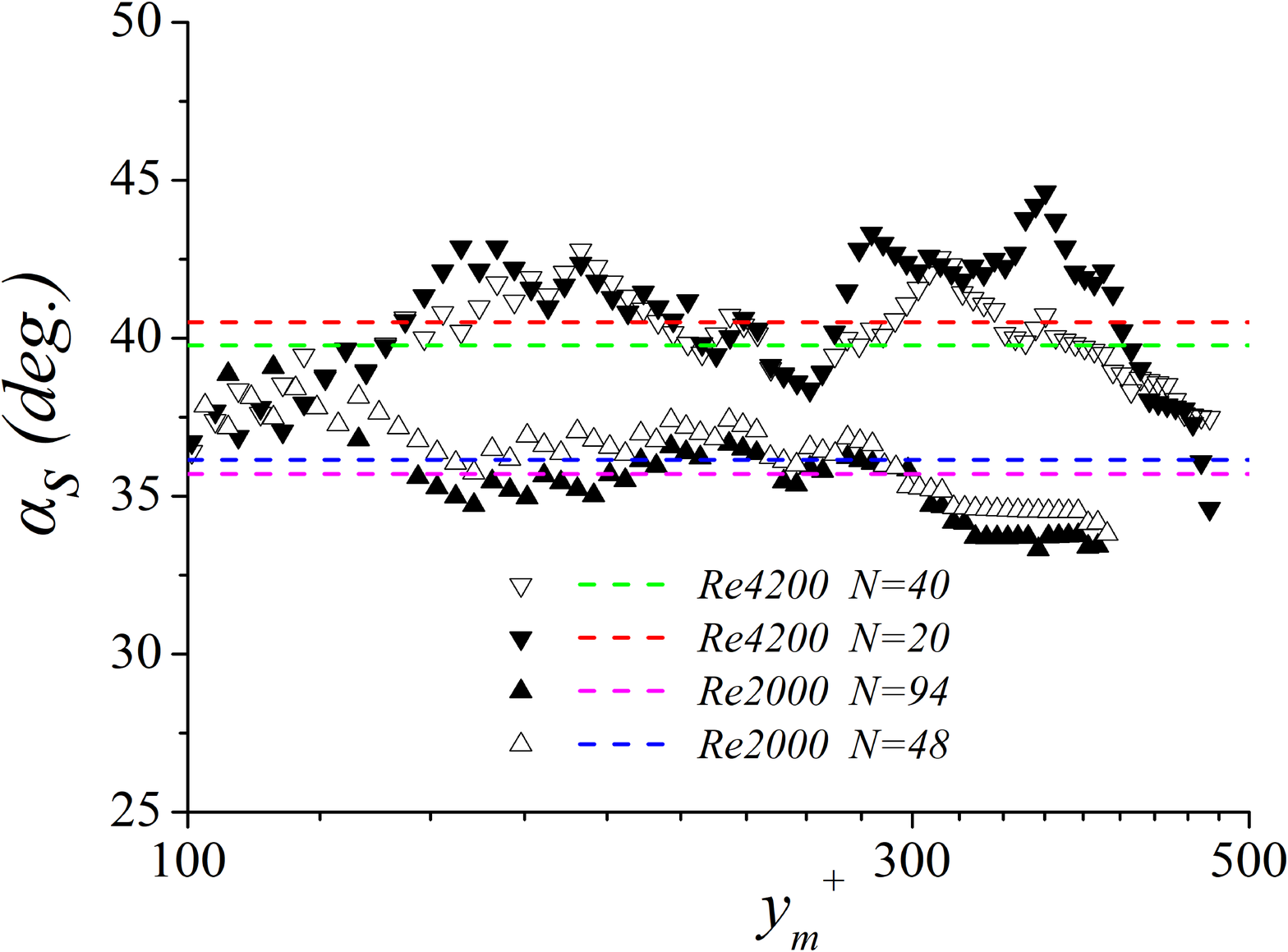} 
}
\caption{$\alpha_s$ as functions of $y_m^+$ for the cases Re2000 and Re4200 with different $N_F$. The dashed lines denote the mean value of $\alpha_s$ in the logarithmic region.}
\label{fig:less} 
\end{figure}

The influences of the number of instantaneous flow fields for accumulating statistics are examined. Fig.~\ref{fig:less} shows the  effect of $N_F$ on the statistic $\alpha_s$ for the cases Re2000 and Re4200.
Alteration of the statistical samples mainly affects the relative standard deviations (RSD) of the results. To be specific, when $N_F$ increases from $48$ to $94$, RSD decreases from $3.9\% $ to $3.3\%$ for Re2000; but for Re4200, RSD decreases from $6.5\% $ to $3.7\%$ when $N_F$ increases from $20$ to $40$. Given the fact that the case Re4200 has limited domain size, raising $N_F$ can effectively reduce the wiggles in the outputs. Nevertheless, the mean value of $\alpha_s$ in the logarithmic region seems to be insensitive to $N_F$.

\bibliographystyle{jfm}
\bibliography{eddy2}

\end{document}